\documentclass[12pt]{article}
\usepackage{epsfig}
\usepackage{graphics}
\usepackage{times}
\begin{document}

{\bf Differential rotation of the chromosphere in the He I absorption line} \\

K. J. Li$^{1, 3, 4}$, J. C.  Xu$^{1}$, J. L.  Xie$^{1,4}$,  W. Feng$^{2}$  \\
$^{1}$Yunnan Observatories, CAS, Kunming 650011, China \\
$^{2}$Research Center of Analysis and Measurement, Kunming University of Science and Technology, Kunming 650093, China \\
$^{3}$Center for Astronomical Mega-Science, Chinese Academy of Sciences, Beijing 100012, China  \\
$^{4}$Key Laboratory of Solar Activity, National Astronomical Observatories, CAS, Beijing 100012, China \\

\date{}
\baselineskip24pt

\begin{abstract}
Differential rotation is the basis of the solar dynamo theory.
Synoptic maps of He I intensity  from Carrington rotations 2032 to 2135 are utilized to investigate the differential rotation of the solar chromosphere in the He I  absorption line. The chromosphere is surprisingly found to rotate faster than the photosphere below it. The anomalous  heating of the chromosphere and corona has been a big problem in modern astronomy.
It is speculated that the small-scale magnetic elements with magnetic flux in the range of $(2.9 - 32.0)\times 10^{18}$ Mx which are anchored in the leptocline,
heat the quiet chromosphere to present the anomalous  temperature increase, causing it to rotate at the same rate as the leptocline.
The differential of rotation rate in the chromosphere is found to be strengthened by strong magnetic fields, but in stark contrast, at the photosphere strong magnetic fields repress the differential of rotation rate. A plausible explanation is given for these findings.  \\
{\bf keywords} Sun: rotation-- Sun: chromosphere-- Sun: activity
\end{abstract}

\section{Introduction}
The sun is magnetic, and solar activity and variation  are both  related to  solar magnetic fields (Fang et al. 2008).
Solar differential rotation is a classical research topic in solar physics due to its close relationship  with  solar magnetic fields (Babcock 1961), and measurements
of differential rotation in the solar atmosphere and especially in the solar interior can provide some clues for solar dynamo theory.
Two kinds of  methods have been utilized to measure the differential rotation in the solar atmosphere. One is direct observation measurements, such as  positioning and tracking  of solar long-life structures (tracers) and Doppler spectrum observations, and the other is the periodicity analyses of  long-term solar activity indexes modulated by solar rotation (Howard 1984; Javaraiah 2003; Chandra, Vats, $\&$ Iyer 2010; Vats $\&$ Chandra 2011; Bhatt et al. 2017; Li et al. 2019; Xiang, Ning, $\&$ Li 2020). Additionally, helioseismic measurements have recently used to detect interior motions of the sun, such as the inner zonal and meridional flows, and the flows are the foundation for understanding the differential  rotation of the solar atmosphere (Howe 2009). Sunspots are the most classical tracer to measure the rotation rate of the solar atmosphere. Generally  magnetic elements in quiet regions rotate  faster than sunspots in active regions (Howard 1984; Xiang et al. 2014; Xu $\&$ Gao 2016), and they both rotate faster than the photospheric atmosphere (Stenflo 1989;
Beck 2000; Lamb 2017; Li et al. 2019; Bertello et al. 2020). Recently  Sudar et al. (2016) and Li et al. (2019) found that the corona anomalously rotates faster than the photosphere below it, although temperature and rotation rate should theoretically decrease (actually increase) from the bottom of the atmosphere, the photosphere to a higher layer of the solar atmosphere, because thermal energy and material of the solar atmosphere flow outward from the photosphere. Combined with the abnormal heating of the upper solar atmosphere is abnormally the rapid rotation of the coronal atmosphere.

The chromosphere is daily observed in the He I $10830\AA$ line  at  National Solar Observatory (NSO)/Kitt Peak (Livingston et al. 1976).  In this study, synoptical maps observed in the line at NSO are utilized to determine the rotation rate of the chromosphere.

\section{Differential rotation of the solar quiet chromosphere}
The solar chromosphere had been routinely surveyed in the He I $10830\AA$  line by the Vacuum Telescope  at NSO/Kitt Peak  from July 2005 to March 2013 (Livingston et al. 1976; Harvey $\&$ Livingston 1994), and correspondingly, synoptic maps of  He I intensity ($I_{He}$) from Carrington rotations 2032 to 2135 were obtained. They can be downloaded from the NSO's web site: ftp://nispdata.nso.edu/kpvt/synoptic/, and  the web site, https://solis.nso.edu/0/vsm/aboutmaps.html introduced how to create a synoptic map from daily observations.
Each synoptic map of  He I intensity was measured at 360 equidistant longitudes from $1^{\circ}$ to $360^{\circ}$ and  180 latitudes ($\varphi$s), and these measurement latitudes have 180 equal steps in the sine of the latitudes ($sin(\varphi)$) ranging from -1 (the solar south pole)  to +1  (the north pole).
Here Figure 1 shows all synoptic maps and an individual map, the synoptic map at Carrington rotation 2057  as an example.  The unit of He I line intensity in these synoptic maps is arbitrary but constant.

The He I line is in absorption at the solar disk (Fleck 1994; Harvey 1994; Brajsa et al. 1996).
Observations of the quiet chromosphere disk are unique in the He I line, that is, structures with strong magnetic field, such as general network bright points,
bright calcium and hydrogen flocculi are dark in the line, displaying absorption features with relatively low intensities, instead of bright structures in the chromosphere  generally observed by  other chromosphere lines, such as $H_{\alpha}$, Ca II, Mg II, and so on
(Harvey $\&$  Sheeley 1977;  Golub et al. 1989;  Fleck et al. 1994;   Brajsa et al. 1996).
Therefore the quiet chromosphere can be approximated through removing low-intensity values, and next in order to achieve this goal,
 we will count the intensity distribution of these synoptic maps to evaluate the proportion of low intensity values in the entire data.
Resultantly, Figure 2 shows the distribution of the original data points of $I_{He}$ intensity in the range of  $-200  \sim 100$,
and those points beyond the range account only for $ 0.54\%$ of the total points.  We don't know why
the range of $I_{He}$ intensity is very wide, from the minimum of -2398.1 to the maximum of 3034.9.
When $I_{He}$ is less respectively than -50, -30, and -10, the ratio of the number of data points to the total is $4.70\%$, $8.31\%$,  and $17.35\%$, correspondingly.
When $I_{He}$s are in the range of $-200 \sim 100$, and meanwhile just those latitudes which are not larger than $30^{\circ}$  are considered,
the distribution of  $I_{He}$s is counted up,  which is shown in the figure as well, and
then  when $I_{He}$ is less respectively
than -50, -30, and -10, the ratio of the number of data points to the total is is $5.49\%$, $8.57\%$, and $17.84\%$, correspondingly.  Therefore active regions of sunspots with strong magnetic fields are basically excluded, if those $I_{He}$s which are less than -10 are not considered, and then the quiet chromosphere can be approximated.

Because dark structures in the He I chromosphere are usually observed/measured in more than one day, we can derive their rotation rates.
The classical autocorrelation analysis method for a time series is used to determine the rotation period of the series (Xu $\&$ Gao 2016; Li $\&$ Feng 2019).
For a time series $X_{i}$ (i=1, 2,..., N), its autocorrelation coefficients is 1 when there is no relative shift, where N  is the number of data points of the series.
First, the series is one-point shifted with respect to itself, and the unpaired endpoints of the two involved series are deleted. At this time, one series is $X_{i}$ (i=1, 2, ..., N-1), the other is $X_{i}$ (i=2, 3, ..., N), and the correlation coefficient ($CC$) can be obtained for these two series.
Next, the original series is two-point shifted with respect to itself, and the unpaired endpoints of the two series are deleted.
At this time, one series is $X_{i}$ (i=1, 2, ..., N-2), the other is $X_{i}$ (i=3, 4, ..., N), and a new  correlation coefficient  can be obtained then.
The above process is repeated again and again, until relative phase shift is 500 data points.
If there are no observations of $I_{He}$s for a Carrington rotation available, $I_{He}$s for the Carrington rotation are given to a very large negative value.
When the negative value appears in the calculation of a correlation coefficient ($CC$), it and its paired value are ignored, and just the rest data are involved in the calculation.
For a time series at a certain latitude, $CC$ is calculated to vary with relative phase shift of the series vs itself, and the shift which corresponds to the local maximum $CC$ around shifts of 27 days is regarded as the rotation period of the series. As an example Figure 3 shows the calculation results ($CC$ varying with relative phase shift) of 6 time series.
Finally, 180 rotation periods are obtained at the corresponding 180 measurement  latitudes, whose calculation results are given in an animation.
In Figure 3, CR is the length of a Carrington synodic rotation period, and  1 CR = 27.275 days, therefore the resolution of relative phase shifts is $27.275/360\approx 0.076$ (days). After the synodic rotation period ($P$) of a series being obtained, its rotation rate can be then gotten to be $P\times 13.199$ (in degrees/day, Shi $\&$ Xie 2013; Deng et al. 2020), which is shown in Panel I of Figure 4, where 13.199 degrees/day is the rotation rate corresponding to the Carrington synodic period. Meanwhile Panel I of Figure 5 shows $CC$ corresponding to each obtained rotation period, and  all $CC$s in the panel are significant at the $99\%$ confidence level.

Differential rotation rate ($\Omega (\varphi)$) is generally expressed as $$\Omega (\varphi)=A + B sin^{2}(\varphi) + C sin^{4}(\varphi).$$
The expression is used to fit those rotation rates whose latitudes are not higher respectively than $30^{\circ}$, $35^{\circ}$, and $40^{\circ}$.
Resultantly, $CC$s for the fitting lines are 0.946, 0.963, and 0.938 in turn, which are all significant at the $99\%$ confidence level.

Similarly, the autocorrelation analysis method is also used to determine rotation periods of those $I_{He}$s which are not less respectively than -50, -30,  and -10, and
resultantly, Panels II, III, and IV of  Figure 4 show the obtained  rotation rate at each measurement latitude  in turn. Correspondingly, Panels II, III, and IV of  Figure 5  display $CC$ at each measurement latitude  in turn.
For the sake of brevity, the rotation rates determined from $I_{He}$s which are not less respectively than -50, -30, and -10 are called  RR II, RR III,  and RR IV in turn, and those determined from the complete original data are called RR I.

The expression is used to fit those rotation rates whose latitudes are not higher respectively than $30^{\circ}$, $35^{\circ}$, and $40^{\circ}$, and the obtained fitting lines are shown in Figure 4. These fitting lines are all highly significant. The fitting lines respectively for RR I to RR IV at latitudes which are not higher than $30^{\circ}$ are put together at the left top panel of Figure 6.
Figure 6 also shows the fitting lines respectively to RR I to RR IV at latitudes which are not higher than $35^{\circ}$ at its left bottom panel, and at the right panel of the figure displayed are the fitting lines respectively to RR I to RR IV at latitudes which are not higher than $40^{\circ}$.
As Figure 6 displays, rotation  slightly becomes fast at low latitudes after eliminating the influence of strong magnetic fields of active regions, and this result is hardly affected by the low-latitude range   considered.
The Doppler rotation rate of the photosphere is also drawn in Figure 6, which was given by  Snodgrass et al. (1984). The Doppler rotation rate is less than
the rotation rate of the quiet chromosphere, and the (quiet) chromosphere rotates faster than the photosphere overall.

RR I whose latitudes are not higher respectively than $60^{\circ}$, $70^{\circ}$, and $80^{\circ}$ are fitted with the following expression of rotation rate, $\Omega (\varphi)=A + B sin^{2}(\varphi) + C sin^{4}(\varphi) + D sin^{6}(\varphi)$, and resultantly, the obtained fitting lines are shown in Panel I of Figure 7. These fitting lines are all highly significant. Due to low rotation rates around latitudes of about $50^{\circ}$, the fitting line to the latitudes which are not higher  than $60^{\circ}$ does not show a monotonous increase from low  to high latitudes, but the other two fitting lines do.

Similarly, the expression is used to respectively fit RR II, RR III, and  RR IV, at latitudes which are not higher respectively than $60^{\circ}$, $70^{\circ}$, and $80^{\circ}$, and resultantly, Panels II, III, and IV of Figure 7 correspondingly display the fitting lines in turn. All these fitting lines show a clear decrease trend of rotation rate from low  to high latitudes. Therefore for the quiet chromosphere, rotation rate decreases from the equator to high latitudes of about $70^{\circ} \sim 80^{\circ}$ on the whole, and  at latitudes of about $40^{\circ}$ onwards, rotation rate fluctuates greatly.

\section{Conclusions and Discussion}
Synoptic maps of He I intensity, which are measured in the interval of Jul. 2005 to Mar. 2013, are used to determine the rotation rate of the full-disk chromosphere observed in the He I $10830\AA$ absorption line. In the two cases of subtracting and not subtracting large magnetic-field values in active regions  from the original data, the chromosphere is surprisingly found to rotate faster than the photosphere beneath it.
Energy and material flow from the solar interior, through the photosphere to the chromosphere.
Since the rotation of the chromosphere is  obviously different from the photosphere rotation, the driving force for the chromosphere to rotate in this way is certainly from the solar interior, not from the photosphere.

Helioseismic measurements indicate that,
at and just at the depths (the so-called leptocline) of $0.99 R_{\odot}$ to near $1 R_{\odot}$ of the solar interior, the long-term variation of the solar seismic radius is in anti-phase with the Schwabe cycle, and the in-situ magnetic fields are  generated (Lefebvre et al. 2009), here $R_{\odot}$ is the solar radius.
The quiet chromosphere is found to be also in anti-phase with the Schwabe cycle (Li et al. 2020). Therefore, the driving force for the quiet chromosphere  to rotate in such a way is believed to come from the leptocline in all probability, since the two present the same cycle phase.
The small-scale magnetic elements ($SME$s) whose magnetic flux spans $(2.9 - 32.0)\times 10^{18}$ Mx are in anti-phase with the Schwabe cycle and distributed all over the solar disk (Jin et al. 2011; Jin $\&$ Wang 2012). They are thought to be the ``in-situ magnetic fields" generated  in the leptocline, due to their same cycle phase.
It is these small-scale magnetic elements that heat the chromosphere  to lead to its formation, so that its rotation is traced by the rotation of the leptocline. Therefore, the quiet chromosphere, $SME$s, and the leptocline are all in anti-phase with the solar cycle.
Temperature changes from the outward decreasing in the photosphere to the outward increasing in the chromosphere.
Studies of modern observations, theory, and statistics have showed that the quiet chromosphere is heated mainly by  small-scale magnetic activities (Parker 1991; De Pontieu et al. 2017; Li et al. 2018), supporting the primary role of the magnetic elements in the heating and formation of the quiet chromosphere. The magnetic elements are anchored in the solar interior, connecting the chromosphere to the leptocline and transmitting the long-term evolution characteristics of the leptocline to the quiet chromosphere.
Small-scale magnetic activities not only cause the quiet chromosphere to be abnormally heated, but also cause it to abnormally rotate faster than the photosphere plasma.

The rotation rate of small-scale magnetic elements is generally larger than that of sunspots, and they both are larger than that of plasma in the photosphere (Stenflo 1989; Howard 1996; Xiang et al. 2014; Xu $\&$ Gao 2016; Lamb 2017). This is the reason why the rotation rate of the quiet chromosphere is found here to be larger than that of the photosphere.

When  strong magnetic fields are not eliminated, the rotation rate of the chromosphere decreases from the equator to middle latitudes faster than that when they are eliminated, and further,
the more data points of strong magnetic fields are eliminated, the slower the rotation rate of the chromosphere decreases from the equator to middle latitudes.
This means that strong magnetic fields should strengthen the differential of rotation rate.  However in sharp contrast,  strong magnetic fields are found to repress the differential of rotation rate in the photosphere (Brajsa et al. 2006; Li et al. 2103).
Sunspots generally appear at low latitudes ($8^{\circ} \sim 35^{\circ}$), and their rotation rates are shown in Figure 6, which come from  Howard (1996).
As the figure shows, the rotation rate of sunspots is lower than that of the quiet chromosphere at the low latitudes, and thus  the appearance of sunspots at a certain latitude should increase the difference of rotation rate between  the latitude and  the equator. Therefore, the differential of rotation rate in the chromosphere is strengthened  by strong magnetic fields.  The rotation rate of sunspots is larger than that of photosphere plasma, and thus the appearance of sunspots at a certain latitude should decrease the difference of rotation rate between the equator and the latitude. Therefore, the differential of rotation rate in the photosphere is repressed by strong magnetic fields.

Like an ``intruder", active regions of sunspots form an invasion zone into the quiet chromosphere, namely the so-called  ``butterfly diagram",  and the quiet chromosphere superimposed with the butterfly pattern is the observed chromosphere (Li et al 2020), as Figure 1 displays here. Similarly,
active regions of sunspots, which take along the general rotation characteristics of the tachocline beneath the lepocline, invade the background flow field of the leptocline, forming a low-speed band with the ``butterfly diagram" migration, which is the observed zonal flow at the leptocline. At the same time, a high-speed band with the ``butterfly diagram" migration (the so-called torsional oscillations) appears in the photosphere (Howard $\&$ LaBonte 1980). Therefore in the photosphere, the rotation rate of sunspots is larger than that of the photosphere plasma, and the differential of rotation rate is  depressed by the rotation of sunspots. All in all, rotation activities in the solar atmosphere are highly related to the magnetic activities anchored in the solar interior, and $SME$s are inferred to originate from the leptocline, while sunspots should come from deeper layers of the interior, probably from the tachocline.

Among the used  data of 104 rotations, 7 rotations are missed (Carrington rotations 2041, 2042, and 2091 to 2095), accounting for $6.7\%$ of the total.
When the missing data are filled with adjacent Carrington rotations, the obtained result is almost the same as that when missing rotations are left out of the analysis. Therefore, such the  small data loss does not have a significant impact on the obtained conclusion.

Synoptic maps of He I intensity are affected by both the projection around the solar poles and the limb darkening, so rotation rates are of lower reliability at high latitudes. This is inferred to be one of the reasons for  a large scatter  of the chromosphere rotation rates at high latitudes.  Zhang, Zirin,  $\&$ Marquette (1997) also found a great fluctuation in the rotation rates at  high latitudes of over $55^{\circ}$. At high latitudes, the magnetic fields are largely  unipolar (Stenflo 1989), and their dependence on time is not explicit (Zhang, Zirin,  $\&$ Marquette 1997), that is, it is relatively difficult to distinguish the appearance and disappearance of a single magnetic field in the time series of magnetic fields, so the determination of the rotation rate of magnetic fields is of greatly uncertainty. This is the main reason why the autocorrelation profile of a time series of magnetic fields is wider at high latitudes than that at low latitudes, whose peak overlays a slowly decaying background (Stenflo 1989), and a reason why a large scatter  of the chromosphere rotation rates appears here at high latitudes.
In this study, latitudes higher than $80^{\circ}$ are not considered, and three latitude bands ($0 - 60^{\circ}$, $0 - 70^{\circ}$, and $0 - 80^{\circ}$) are separately considered, in order to somewhat present the influence of these factors on the obtained results. In these three cases, rotation rate for the quiet chromosphere is found to present a decrease trend from the equator to high latitudes as well, but this finding does not have high credibility and need be demonstrated by the precise observations of solar polar regions in the future.

{\bf acknowledgments:}
We thank the anonymous referee for careful reading of the manuscript and constructive comments which improved the original version of the manuscript.
The full-disk synoptic maps of He I intensity are courtesy to be publicly downloaded from the NSO's web site.
The authors would like to express their deep thanks to the staffs of the web site.
This work is supported by the
National Natural Science Foundation of China (11973085, 11903077, 11803086, 11703085, and 11633008), the Yunnan Ten-Thousand Talents Plan (the Yunling-Scholar Project), the national project for large scale scientific facilities (2019YFA0405001),  the CAS ``Light of West China" Program, and  the Collaborating Research Program of CAS Key Laboratory of Solar Activity (KLSA201912,KLSA202012), and the Chinese Academy of Sciences.

\clearpage

\begin{figure*}
\begin{center}
\centerline{\includegraphics[width=1.05 \textwidth]{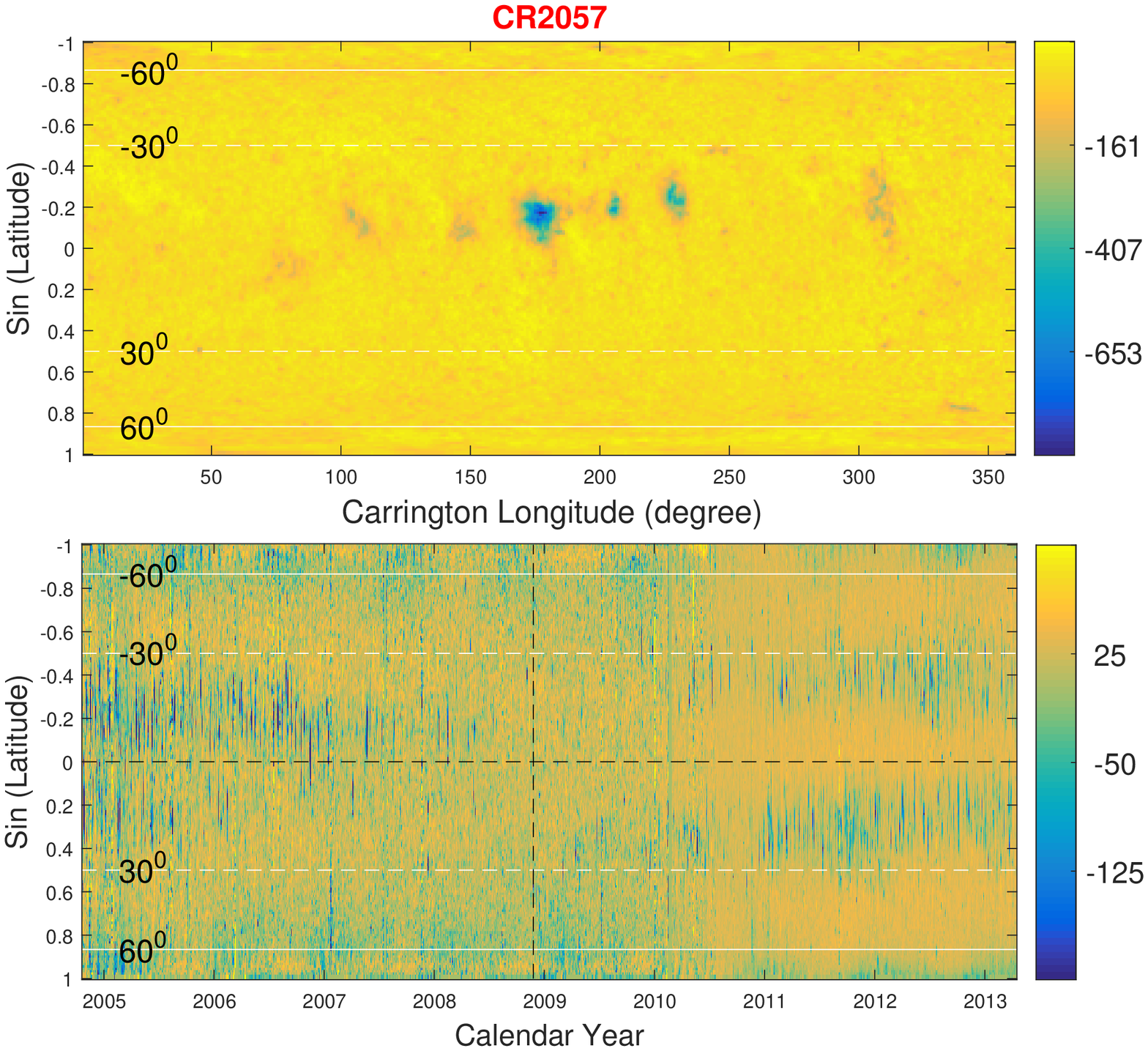}}
\caption{Top panel: the synoptic map of He I  intensity at Carrington rotation  2057. Bottom panel: the time-latitude diagram of He I intensity in the interval of Jul. 2005 to Mar. 2013.
The white dashed (solid) lines show the latitudes of $\pm 30^{\circ}$ ($\pm 60^{\circ}$), and the vertical dashed line shows the minimum time of cycle 24.
 }\label{}
\end{center}
\end{figure*}

\begin{figure*}
\begin{center}
\centerline{\includegraphics[width=1.05 \textwidth]{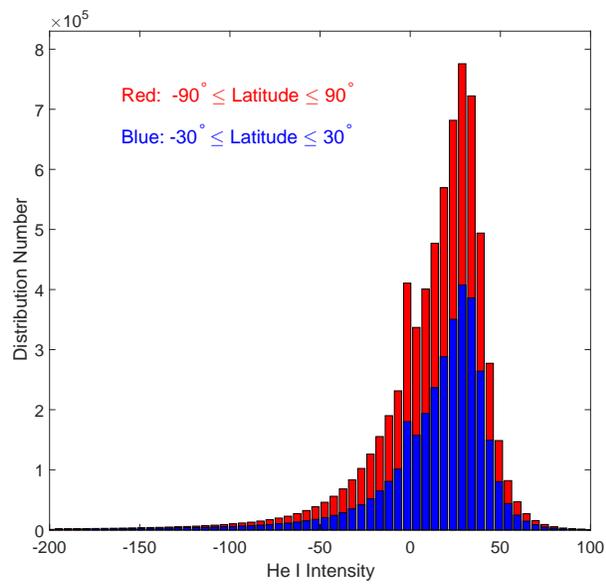}}
\caption{The distribution of He I  intensity ($I_{He}$) in the range of that $-200 \leq I_{He} \leq 100$ and meanwhile respectively at the full disk (the red histogram) and  latitudes (the blue histogram) of $-30^{\circ}$ (the southern hemisphere) to $30^{\circ}$ (the northern hemisphere).
 }\label{}
\end{center}
\end{figure*}

\begin{figure*}
\begin{center}
\centerline{\includegraphics[width=1.05 \textwidth]{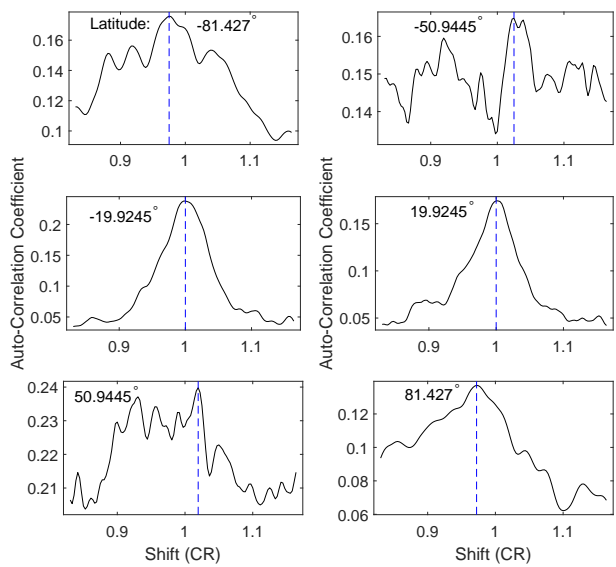}}
\caption{Auto-correlation coefficient ($CC$) of a time series of $I_{He}$ at a certain latitude, varying with its relative phase shift. 6 panels correspond to 6 time series of $I_{He}$ whose latitudes are indicated at the top of each panel. The blue vertical dashed line in each panel shows the local maximum $CC$, whose abscissa shift is the synodic rotation period in CRs, and 1CR=27.28 days. (An animation of this figure is available.)
 }\label{}
\end{center}
\end{figure*}

\begin{figure*}
\begin{center}
\centerline{\includegraphics[width=1.05 \textwidth]{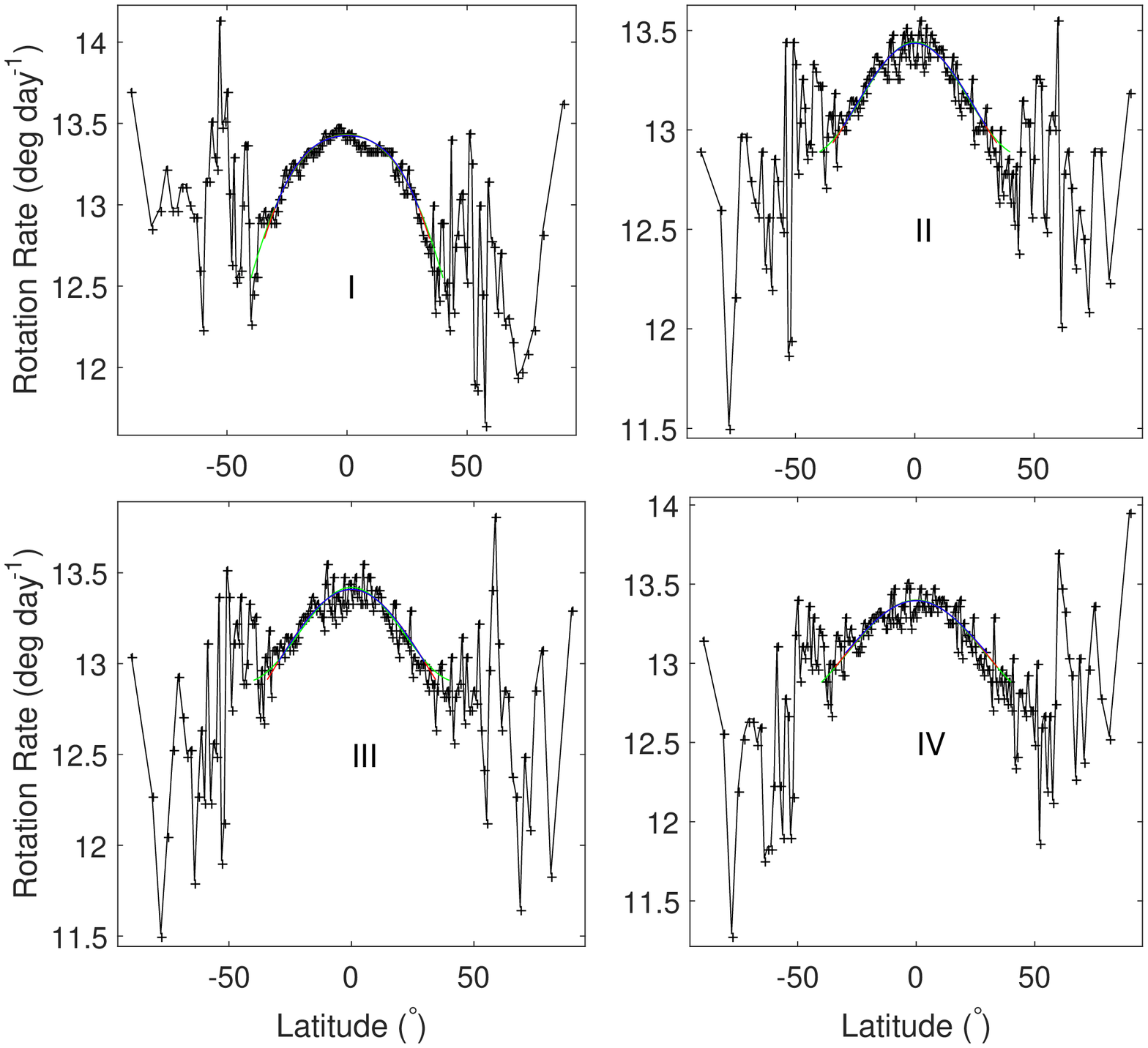}}
\caption{Top left panel (Panel I): rotation rate (crosses in the black line) at each measurement latitude. The blue, red, and green lines are the fitting lines of those rotation rates whose latitudes are not greater than $30^{\circ}$, $35^{\circ}$, and $40^{\circ}$, correspondingly.
Top right panel (Panel II): the same as Panel I, but those $I_{He}$s which are not less  than -50 are considered.
Bottom left panel (Panel III): the same as Panel I, but those $I_{He}$s which are not less  than -30 are considered.
Bottom right panel (Panel IV): the same as Panel I, but those $I_{He}$s which are not less  than -10 are considered.
 }\label{}
\end{center}
\end{figure*}

\begin{figure*}
\begin{center}
\centerline{\includegraphics[width=1.05 \textwidth]{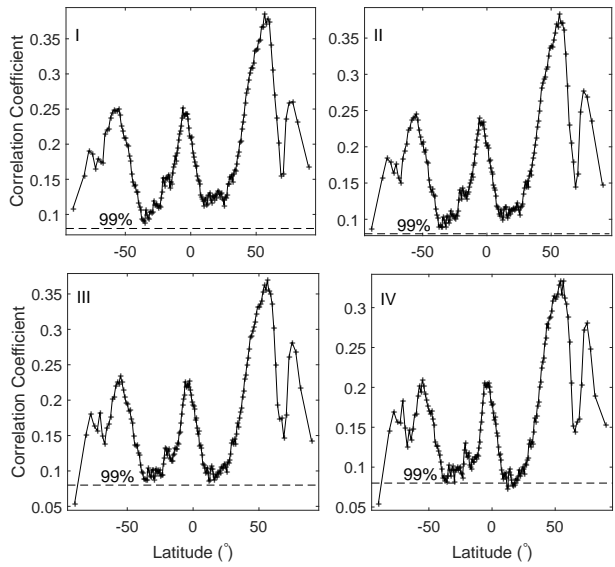}}
\caption{Top left panel (Panel I): auto-correlation coefficient (crosses in the solid line) corresponding to the determined  rotation period for a series of $I_{He}$ at a measurement latitude. The dashed line shows the $99\%$  confidence level.
Top right panel: the same as Panel I, but those $I_{He}$s which are not less than -50 are considered.
Bottom left panel: the same as Panel I, but those $I_{He}$s which are not less than -30 are considered.
Bottom right panel: the same as Panel I, but those $I_{He}$s which are not less than -10 are considered.
  }\label{}
\end{center}
\end{figure*}

\begin{figure*}
\begin{center}
\centerline{\includegraphics[width=1.05 \textwidth]{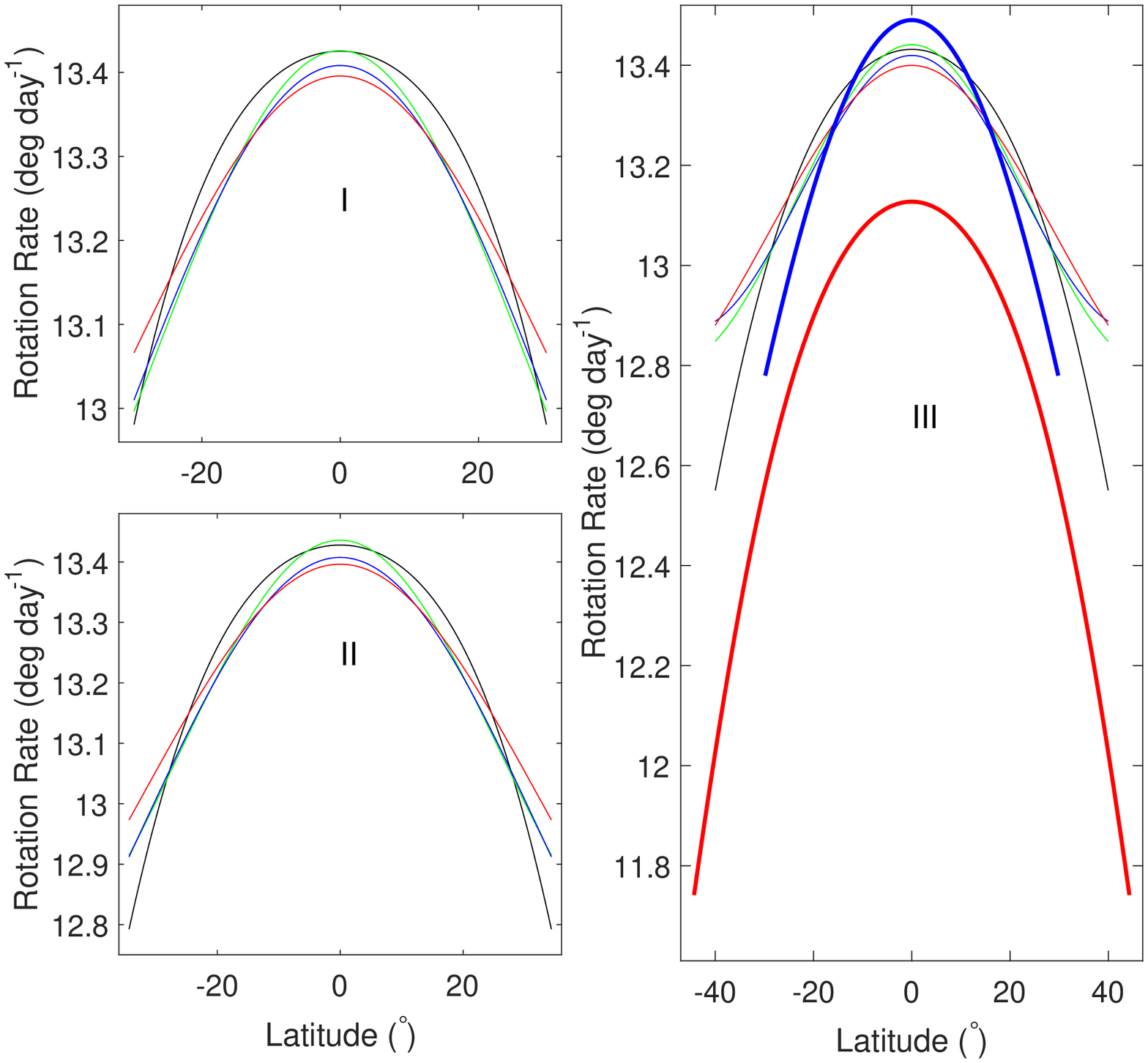}}
\caption{Left top panel:  the fitting lines respectively to RR I (the black solid line),  RR II (the green line), RR III (the blue thin line), and  RR IV (the red thin line),  at latitudes which are not greater than $30^{\circ}$.
Left bottom panel: the same as the left top panel, but  at latitudes which are not greater than $35^{\circ}$, instead of $30^{\circ}$.
Right panel: the same as the left top panel, but  at latitudes which are not greater than $40^{\circ}$, instead of $30^{\circ}$.
In the right panel, the red thick line shows the rotation rate of the photosphere, which is determined in spectrum method by  Snodgrass et al. (1984); and the blue thick lines show the rotation rate of sunspots, which is given by  Howard (1996).
 }\label{}
\end{center}
\end{figure*}

\begin{figure*}
\begin{center}
\centerline{\includegraphics[width=1.05 \textwidth]{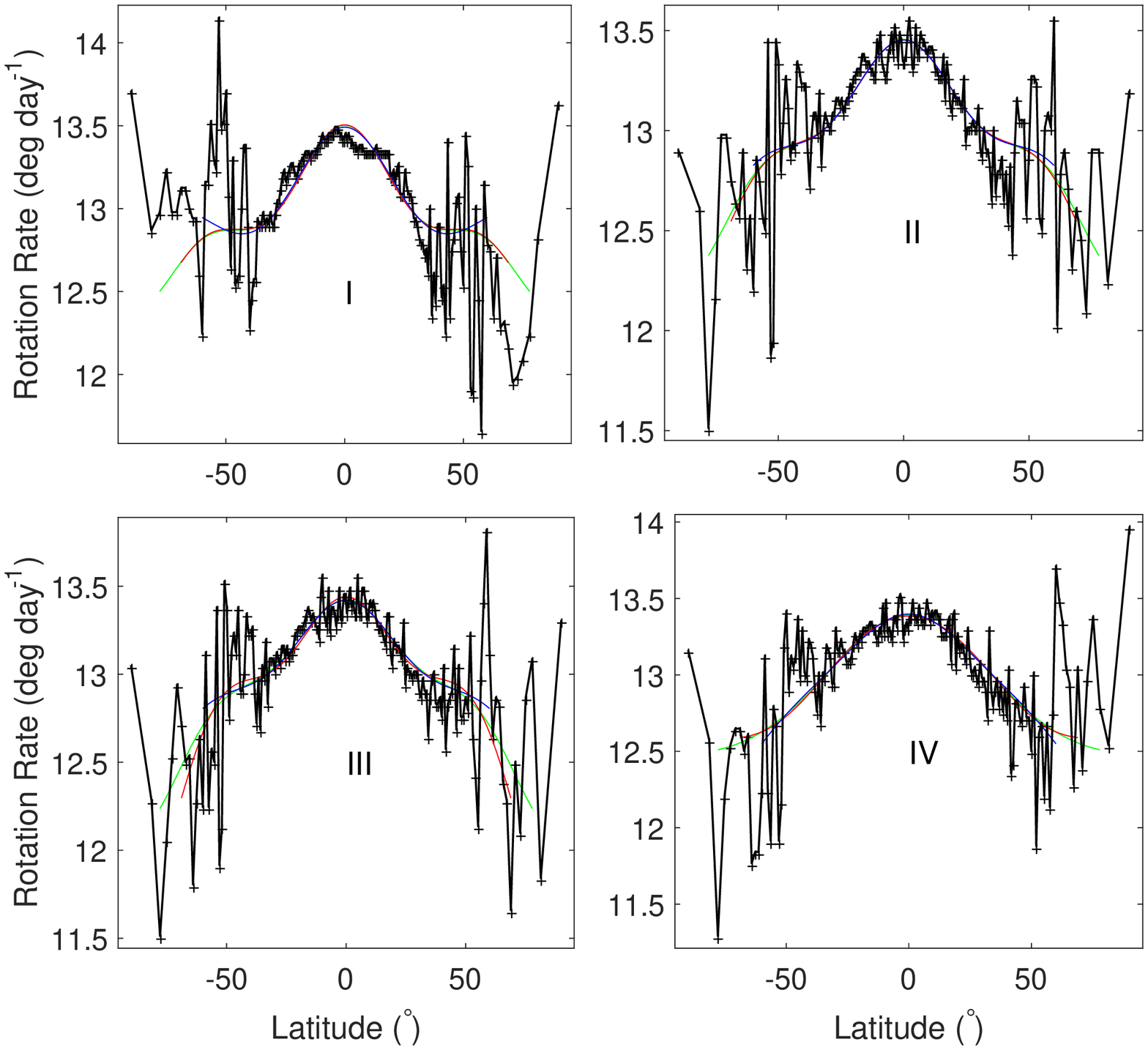}}
\caption{Top left panel (Panel A): rotation rate (crosses in the black line) at each measurement latitude. The blue, red, and green lines are the fitting lines of those rotation rates whose latitudes are not greater than $60^{\circ}$, $70^{\circ}$, and $80^{\circ}$, correspondingly.
Top right panel (Panel B): the same as Panel A, but those $I_{He}$s which are not less  than -50 are considered.
Bottom left panel (Panel C): the same as Panel A, but those $I_{He}$s which are not less  than -30 are considered.
Bottom right panel (Panel D): the same as Panel A, but those $I_{He}$s which are not less  than -10 are considered.
}\label{}
\end{center}
\end{figure*}

\end{document}